\documentclass[runningheads]{llncs}

\usepackage[T1]{fontenc}
\def\doi#1{\href{https://doi.org/\detokenize{#1}}{\url{https://doi.org/\detokenize{#1}}}}
\usepackage{subfigure}
\usepackage{graphicx}
\usepackage{amsmath}
\usepackage{newtxmath}
\usepackage{amsfonts}
\usepackage{multirow}
\usepackage{tabularx}
\usepackage{ulem}
\usepackage{bm}
\usepackage[dvipsnames]{xcolor}
\usepackage{threeparttable}
\usepackage{amsmath}
\usepackage[pagebackref=true,breaklinks=true,colorlinks,bookmarks=false]{hyperref}

\usepackage{booktabs}

\usepackage{color}

\usepackage{listings}
\begin{document}
\title{A CT Image Denoising Method with Residual Encoder-Decoder Network}

\author{Helena Shawn \inst{1}
\and Thompson Chyrikov\inst{1}
\and Jacob Lanet\inst{1}
\and Lam-chi Chen\inst{1,2}
\and Jim Zhao \inst{1, 2}
\and Christina Chajo \inst{1} \thanks{Christina Chajo is corresponding author.}}
\authorrunning{H. Shawn et al.}
%
\institute{University of Texas Medical Branch in Galveston, USA\\ 
National University of Science and Technology, TW
}
\maketitle              
\begin{abstract}
Utilizing a low-dose CT approach significantly reduces the radiation exposure for patients, yet it introduces challenges, such as increased noise and artifacts in the resultant images, which can hinder accurate medical diagnostics. Traditional methods for noise reduction struggle with preserving image textures due to the complexity of modeling statistical properties directly within the image domain. To address these limitations, this study introduces an enhanced noise-reduction technique centered around an advanced residual encoder-decoder network. By incorporating recursive processing into the foundational network, this method reduces computational complexity and enhances the effectiveness of noise reduction. Furthermore, the introduction of a root-mean-square error and perceptual loss functions aims to retain the integrity of the images' textural details. The enhanced technique also includes optimized tissue segmentation, improving artifact management post-improvement. Validation using the TCGA-COAD clinical dataset demonstrates superior performance in both noise reduction and image quality, as measured by post-denoising PSNR and SSIM, compared to the existing WGAN approach. This advancement in CT image processing offers a practical solution for clinical applications, achieving lower computational demands and faster processing times without compromising image quality.


\end{abstract}

\section{Introduction}
In the realm of medical imaging~\cite{arnaout2021ensemble,CW2017evaluate,iandola2014densenet}, particularly computed tomography (CT), advancements in technology have led to increased reliance on CT scans for diagnostic purposes. These scans, while invaluable for their detailed insights into the human body, have raised concerns due to the high doses of radiation they expose patients to. A solution that has emerged is the development of low-dose CT (LDCT) scanning techniques~\cite{chen2020label,Dong2018Boosting}, which reduce radiation exposure by lowering the X-ray tube current. However, this reduction in radiation dose compromises the signal-to-noise ratio in the projection data, leading to images marred by noise and artifacts when reconstructed using conventional methods like filtered back projection (FBP). The presence of these imperfections in CT images complicates the task of accurate clinical diagnosis, highlighting the critical need for effective methods to enhance the quality of LDCT images.

The endeavor to improve LDCT image quality has given rise to various strategies, including projection domain denoising algorithms, image reconstruction algorithms, and image domain denoising algorithms. Despite these efforts, achieving satisfactory enhancement in CT image quality remains a challenge. Traditional denoising methods are not only computationally intensive and time-consuming but also struggle to model the complex distribution of noise post-processing accurately. As a result, these methods may reduce noise and artifacts at the cost of blurring important edges and details, rendering the denoised CT images less useful for clinical diagnosis.

The advent of deep neural networks (DNNs) has introduced a novel approach to LDCT image denoising, leveraging the networks' robust feature learning and mapping capabilities. DNNs have demonstrated superior performance in terms of reconstruction quality and processing speed compared to traditional methods. Nonetheless, the conventional use of mean square error (MSE) as a loss function in these networks often leads to oversmoothing of edges and loss of detail, failing to preserve image textures that are crucial for human visual perception. This indicates a pressing need for innovative solutions that can effectively balance noise reduction with the preservation of critical image features, thereby meeting the stringent demands of clinical diagnosis.

\section{Related works}
Various techniques used in image processing and computer vision, 
such as image filtering, edge detection, and feature extraction has been proposed. These techniques can be used in image denoising, for example, to identify edges in an image and preserve their structure while removing noise. And we learnt to use different types of filters for different kinds of noises. There have been many research papers published on image denoising, and 
various approaches have been proposed to solve this problem. Some popular methods include wavelet-based denoising~\cite{gomez2010multi}, non-local means denoising~\cite{jeong2006biologically}, sparse representation-based denoising~\cite{yang2018low} and task-oriented denoising~\cite{zhang2021task}.  
Among all the methods, edge enhancing method is a challenging yet crucial approach. Many techniques proposed in previous studies, such as histogram 
equalization, are sensitive to noise and can unintentionally enhance both the image structures and the noise. While some previous methods have been 
introduced for image enhancement, particularly in the field of medical 
imaging~\cite{zhang2022overlooked}, they do not address the issue of noisy images. In fact, these approaches tend to struggle when dealing with images that have low peak signal-to-noise ratio (PSNR) as the noise is often amplified alongside the signal. For example, there exist some other image denoising methods~\cite{yang2018low}, using edge detection to enhance the structure feature of the original image while using different filters to remove the random noise. There is also another line.

\section{Methodology}

\begin{figure}[t]
    \centering
    \includegraphics[width=\columnwidth, clip=true, trim=0 0 0 0]{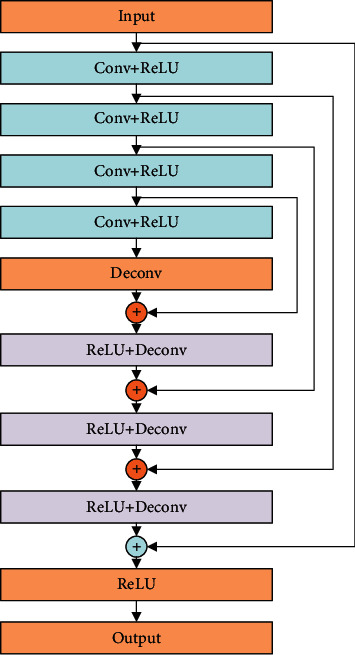}
    \caption{The proposed framework for the proposed residual network for CT image denoising.
}
    \label{fig:fig1}
\end{figure}

The SegNet model, as shown in Fig.~\ref{fig:fig1} comprising a set of convolutional coding layers and mirrored deconvolutional 
decoding layers can achieve better effect in semantic segmentation and thus is selected for the loss 
network. The encoding part uses the visual geometry group (VGG) model with strong generalization 
ability, and the decoding part uses a symmetrical structure to recover the information lost in 
pooling. Besides, the pretrained Caffe model is used to ensure the ability of the loss network to 
extract features.

After determining the loss network, perceptual loss needs to be defined at the semantic feature 
level. The specific steps are as follows: input the fuzzy denoising result \( x - F(x_{detail}) \) and real image \( y \)
initially generated by the front-end network into SegNet. The feature maps of these two are 
extracted from the fixed convolutional layer, and then the Euclidean distance represented by these 
two features is calculated, as shown in the following equation:
\[
L_{Per} = \frac{1}{W_i H_i} \| \phi_i (x - F(x_{detail})) - \phi_i (y) \|_2^2,
\]
where \( W_i \) and \( H_i \) represent the width and height of the selected feature map, respectively, and \( \phi_i \) is 
the feature map extractor.

The joint perceptual loss consists of two parts, namely MSE and perceptual loss. The structural 
model is shown in Figure 3. The method is implemented as follows: first, noisy CT image and real CT 
image \( y \) are input into the denoising network. The difference between these two is compared and 
learned pixel by pixel through the MSE loss function, and the initial denoising result \( x - F(x_{detail}) \) that 
matches pixel \( y \) is obtained. At this time, the CT image after initial denoising is blurry. On this basis, \( y \) 
and \( x - F(x_{detail}) \) are input into the loss network SegNet, respectively. Then the two feature maps \( \phi(x - F(x_{detail})) \) and \( \phi(y) \) are extracted from one of the convolutional layers to define the perceptual loss 
function. The network continues to train by minimizing perceptual loss to learn the difference in 
semantic features of these two images, reconstruct the edge and detail information, and make the 
two images more similar in feature perception. Finally, a clearer CT image denoising result is 
generated.

\section{Experimental Results}

\begin{figure}[t]
    \centering
    \includegraphics[width=\columnwidth, clip=true, trim=0 0 0 0]{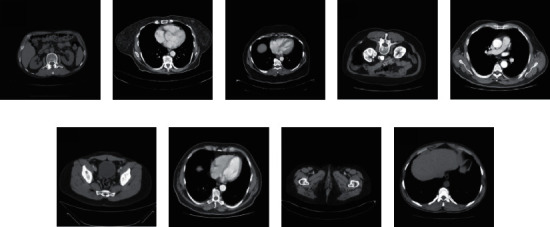}
    \caption{A case study for CT image denoising.
}
    \label{fig:fig2}
\end{figure}

From the TCGA-COAD clinical dataset, nine CT images were chosen at random to serve as evaluation samples. These were distinct and did not coincide with the 200 images earmarked for training purposes, as illustrated in Figure~\ref{fig:fig2}. For comparative analysis, this study utilizes the WGAN algorithm and the foundational RED-CNN algorithm . The noise reduction capabilities of these methods are exemplified in Figure 6, where sample (a) and sample (c) from Figure 5 are used to demonstrate the denoising outcomes.

This paper selects hip joint prosthesis as the test set to evaluate the image quality of the network after training with the number of layers at 4, 8, and 12, respectively. Table~\ref{tab:my_label} shows SSIM value, RMSE value, and PSNR value of images after three network training. As can be seen from Table~\ref{tab:my_label}, when the number of RED network layers is 8, the output image shows a better evaluation index: a larger SSIM value, a smaller RMSE value, and a larger PSNR value, which indicate that the network has a higher performance in image restoration. In summary, when the network model has 8 layers, it can achieve faster convergence speed, higher image quality after correction, and better performance of metal artifact correction.

\begin{table}[h]
\centering
\begin{tabular}{cccc}
\hline
\textbf{Network layers} & \textbf{SSIM} & \textbf{RMSE} & \textbf{PSNR} \\
\hline
4 & 0.9584 & 0.0065 & 68.625 \\
8 & 0.9592 & 0.0062 & 68.634 \\
12 & 0.9576 & 0.0067 & 68.612 \\
\hline
\end{tabular}
\caption{SSIM, RMSE, and PSNR values under RED-CNN with the number of layers at 4, 8, and 12.}
\label{tab:my_label}
\end{table}

\section{Conclusions}

The complexity inherent in characterizing statistical elements within the realm of image domains poses a significant challenge to current methodologies that seek to process reconstructed images. These methods often fail to adequately suppress image noise without sacrificing the intricate details of the image's structure. The advent of deep learning has opened new avenues in the restoration of noise artifacts in low-dose CT (LDCT) images, presenting a promising field of exploration. Addressing the suboptimal noise reduction capabilities of conventional algorithms, the cumbersome nature of complex network models, and the obstacles presented by training procedures, this study introduces an enhanced method for CT image denoising. It leverages a refined RED network approach, aiming to strike a balance between noise reduction and preservation of critical image details.

%
%
%
%
\newpage
\bibliographystyle{splncs04}
\bibliography{refs}
\end{document}